\documentclass[utf8]{frontiersSCNS} 
\usepackage{url,hyperref,lineno,microtype,subcaption}
\usepackage[onehalfspacing]{setspace}
\usepackage{graphicx}
\usepackage{lipsum}
\usepackage{amsmath}
\usepackage{amsfonts}
\usepackage{amssymb}
\usepackage{csquotes}
\usepackage{bm}
\usepackage{pbox}
\usepackage{tabularx}
\usepackage{booktabs}
\usepackage{tikz}

\usepackage{enumitem}
\usepackage[justification=centering]{caption}

\newcolumntype{R}[1]{>{\raggedleft\arraybackslash}p{#1}}
\newcolumntype{L}[1]{>{\raggedright\arraybackslash}p{#1}}
\newcolumntype{C}[1]{>{\centering\arraybackslash}p{#1}}

\newcommand{\pkg}[1]{{\normalfont\fontseries{b}\selectfont #1}}
\let\proglang=\textsf
\let\code=\texttt

\definecolor{myred}{HTML}{A65554}
\definecolor{mygreen}{HTML}{6B8C54}
\definecolor{myblue}{HTML}{2E8AB2}
\definecolor{myyellow}{HTML}{B29440}

\usepackage{blindtext}
\usepackage{scrextend}
\addtokomafont{labelinglabel}{\sffamily}

\usepackage{etoolbox}

\BeforeBeginEnvironment{tabular}{\begin{center}\footnotesize}
\AfterEndEnvironment{tabular}{\end{center}}

\def\keyFont{\fontsize{8}{11}\helveticabold }
\def\firstAuthorLast{Onsj\"o and Sheridan} 
\def\Authors{Mikael Onsj\"o\,$^{1}$ and Paul Sheridan\,$^{2,*}$}
\def\Address{$^{1}$Independent Researcher \\
$^{2}$Department of Active Life Promotion Science, Graduate School of Medicine, Hirosaki University, 5 Zaifu, Hirosaki, Aomori, Japan, 036-8562}
\def\corrAuthor{Paul Sheridan}

\def\corrEmail{paul.sheridan.stats@gmail.com}

\begin{document}
\onecolumn
\firstpage{1}

\title[Theme Enrichment Analysis]{Theme Enrichment Analysis: A Statistical Test for Identifying Significantly Enriched Themes in a List of Stories with an Application to the Star Trek Television Franchise} 

\author[\firstAuthorLast ]{\Authors} 
\address{} 
\correspondance{} 

\extraAuth{}

\maketitle

\begin{abstract}

\section{}
In this paper, we describe how the hypergeometric test can be used to determine whether a given theme of interest occurs in a storyset at a frequency more than would be expected by chance. By a storyset we mean simply a list of stories defined according to a common attribute (e.g., author, movement, period). The test works roughly as follows: Given a background storyset and a sub-storyset of interest, the test determines whether a given theme is over-represented in the sub-storyset, based on comparing the proportions of stories in the sub-storyset and background storyset featuring the theme. A storyset is said to be ``enriched'' for a theme with respect to a particular background storyset, when the theme is identified as being significantly over-represented by the test. Furthermore, we introduce here a toy dataset consisting of 280 manually themed Star Trek television franchise episodes. As a proof of concept, we use the hypergeometric test to analyze the Star Trek stories for enriched themes. The hypergeometric testing approach to theme enrichment analysis is implemented for the Star Trek thematic dataset in the \proglang{R} package \pkg{stoRy}. A related \proglang{R} \pkg{Shiny} web application can be found at \url{https://github.com/theme-ontology/shiny-apps}.

\tiny
 \keyFont{ \section{Keywords:} enrichment analysis, hypergeometric test, over-representation analysis, Star Trek, theme ontology} 
\end{abstract}

\section{Introduction}
A \emph{literary theme}, or \emph{theme} for short, is loosely defined as ``An idea that recurs in or pervades a work of art or literature'' (e.g., a story)~\footnote{Oxford Dictionaries, Oxford University Press (2018). URL: \url{https://en.oxforddictionaries.com/definition/theme/}. [Online; accessed 1-May-2018]}. Themes are often expressible in a single word or short phrase, as is illustrated by such garden-variety themes as ``love'', ``loyalty'', and ``the lust for gold''.~These examples all happen to be value-neutral abstractions. But themes can just as well take the form of morally charged messages, such as ``be wary of strangers'' and ``do not judge prematurely''. The consummate story-maker usually takes pains to imply a theme indirectly, rather than state it explicitly. Sometimes the story-maker is even unconscious of important themes found in their stories. A~typical story will feature multiple themes. In the present work, we distinguish between \emph{central} themes (i.e., themes found to recur throughout a major part of a story or otherwise important to its conclusion) and \emph{peripheral} themes (i.e., briefly featured themes that are not part of the main story narrative). To sum up: the themes of a story furnish the partaker thereof with an executive summary of what the story is truly about.

Enrichment (or over-representation) analysis, which amounts to a class of statistical tests, is well-suited for determining if a given theme of interest is featured more than would be expected by chance in a subset of stories drawn from a larger set of stories. In particular, the hypergeometric test uses the hypergeometric distribution to calculate a \emph{p}-value associated with having drawn at least $k$ successes in $n$ draws, without replacement, from a population of size $N$ that contains exactly $K$ successes. We define a \emph{storyset} to be a set of stories defined by some common attribute (e.g., author, movement, period). In the present context, it is appropriate to think of the population as a \emph{background storyset} (i.e., a set of $N$ stories), and the sub-population as a \emph{test storyset} (i.e., a subset of $n \leq N$ background stories). A test storyset is said to be \emph{enriched} (or \emph{over-represented}) for a given theme when the associated hypergeometric test \emph{p}-value is lower than a predetermined significance level.

The hypergeometric test is routinely used in bioinformatical analyses to identify over-represented biological terms in lists of genes~\citep{Boyle2004, Zheng2008, Huang2009}. This differs from the corresponding state of affairs in literary studies. To our knowledge, the hypergeometric test is yet to be implemented in any of the textual analysis tools in common usage among humanities scholars, including the Stanford Topic Modeling Toolbox~\citep{Ramage2009}, TAPoR~\footnote{TAPoR 3 Discover research tools for studying texts (2018). URL: \url{http://tapor.ca/tools}. [Online; accessed 1-May-2018]}, TOME~\citep{Klein2015}, Word Seer~\citep{Muralidharan2013}, and Voyant Tools~\footnote{Sinclair, S., Rockwell, G. and the Voyant Tools Team, Voyant Tools (2012). URL: \url{https://voyant-tools.org/}. [Online; accessed 1-May-2018]}. The same holds true of computer-assisted qualitative data analysis software that is sometimes used for document analysis in the social sciences~\footnote{McNiff, K., What is Qualitative Research? (2016).  URL: \url{http://www.qsrinternational.com/nvivo/nvivo-community/blog/what-is-qualitative-research}. [Online; accessed 1-May-2018]}. ATLAS.ti~\footnote{ATLAS.ti Scientific Software Development GmbH (1993-2017). ATLAS.ti.}, NVivo~\footnote{QSR International Pty Ltd. (2015). NVivo qualitative data analysis Software Version 11.}, and MAXQDA~\footnote{VERBI Software (1989-2017). MAXQDA, software for qualitative data analysis.} are three such programs that allow users to manage, analyze, and visualize data related to text, audio, and video documents. However, none of these programs implement the said test at the time of writing.

In this paper, we champion the hypergeometric test as a statistically principled approach to theme enrichment analysis. To this end, we introduce a toy dataset consisting of a total of $280$ themed Star Trek television series episodes. We recorded themes for each of the $80$ episodes of \emph{Star Trek: The Original Series} (\emph{TOS}), $22$ episodes of \emph{Star Trek: The Animated Series} (\emph{TAS}), and $178$ episodes of \emph{Star Trek: The Next Generation} (\emph{TNG}); see Supplementary Note~1 for an overview of the Star Trek franchise. What is more, we hierarchically arranged the collected themes into a draft theme~ontology. The Star Trek thematic dataset paves the way for a demonstration of how the hypergeometric testing approach to enriched theme identification is helpful for summing up what makes a storyset unique and for generating speculative hypotheses. We present the results of two case studies as a proof of concept. The first examines how the portrayal of the Klingons changed from that of a tyrannical and expansionist empire in \emph{TOS} to an inward looking warrior culture in \emph{TNG}. The second explores the most significantly enriched themes for each of \emph{TOS}, \emph{TAS}, and \emph{TNG}. In short, we find that \emph{TOS} stands out for its focus on the social issues of the day and by extension on what constitutes a just and flourishing society, \emph{TAS} for novel sci-fi and fantasy concepts especially suited to an animated series, and \emph{TNG} for its comparatively refined treatment of the human condition. Moreover, we show that our findings compare favorably with those obtained using the standard term frequency--inverse document frequency (TF-IDF) approach to enrichment analysis~\citep{Salton1975}, which is commonly implemented in textual analysis software packages. More specifically, we find that the hypergeometric testing and TF-IDF approaches to enrichment analysis, while agreeing in broad outlines, exhibit significant differences in terms of the top themes they identify as being enriched. This is consequential because the hypergeometric test is, by its very definition, the standard according to which heuristics (i.e.,TF-IDF) ought to be evaluated when it comes to answering the question of whether a theme is enriched in a storyset of interest.

The rest of the paper is organized as follows: In Section~\ref{SEC_TO} we introduce our aforementioned draft theme ontology. It is a hierarchically organized theme vocabulary, partitioned into the following four domains: the human condition, society, the pursuit of knowledge, and alternate reality. There are $1535$ unique themes in total. Criteria and guidelines motivating our hierarchical arrangement are discussed. In Section~\ref{SEC_TEA} we explain the hypergeometric testing approach to theme enrichment analysis in full technical detail. In Section~\ref{SEC_TEA_Application} we use the hypergeometric test to identify enriched themes in the two Star Trek case studies outlined above. These results we compare with those obtained using the standard TF-IDF approach to enrichment analysis. We conclude the paper in Section~\ref{SEC_Discussion} with a summary of our main contributions, a discussion of some limitations of our methodology, and go on to describe a handful of possible future directions. Most notably, in terms of limitations, we emphasize that we manually tagged stories with themes, and as a consequence the findings we report inevitably reflect our point of view, and are not fully replicable. An overview of Star Trek film and television series franchise is found in Supplementary Note~1. A list of episodes used in the Klingon theme enrichment case study is found in Supplementary Note~2. The results of the TF-IDF approach to theme enrichment analysis are presented in Supplementary Note~3. The theme enrichment analysis procedure based on the hypergeometric test is implemented in the \proglang{R} package \pkg{stoRy} (version 0.1.1)~\citep{Sheridan2017}, released through CRAN~\footnote{The Comprehensive R Archive Network (2017). URL: \url{https://cran.r-project.org/}. [Online; accessed 1-May-2018]}. The Star Trek thematic data is included in the package. A related \proglang{R} \pkg{Shiny} web application is available for download at the Theme Ontology~\footnote{Theme Ontology (2018). URL: \url{http://www.themeontology.org}. [Online; accessed 1-May-2018]} GitHub repository at~\url{https://github.com/theme-ontology/shiny-apps}.

\section{A Theme Ontology}\label{SEC_TO}

\begin{figure}[!h]
\begin{center}
\includegraphics[width=14cm]{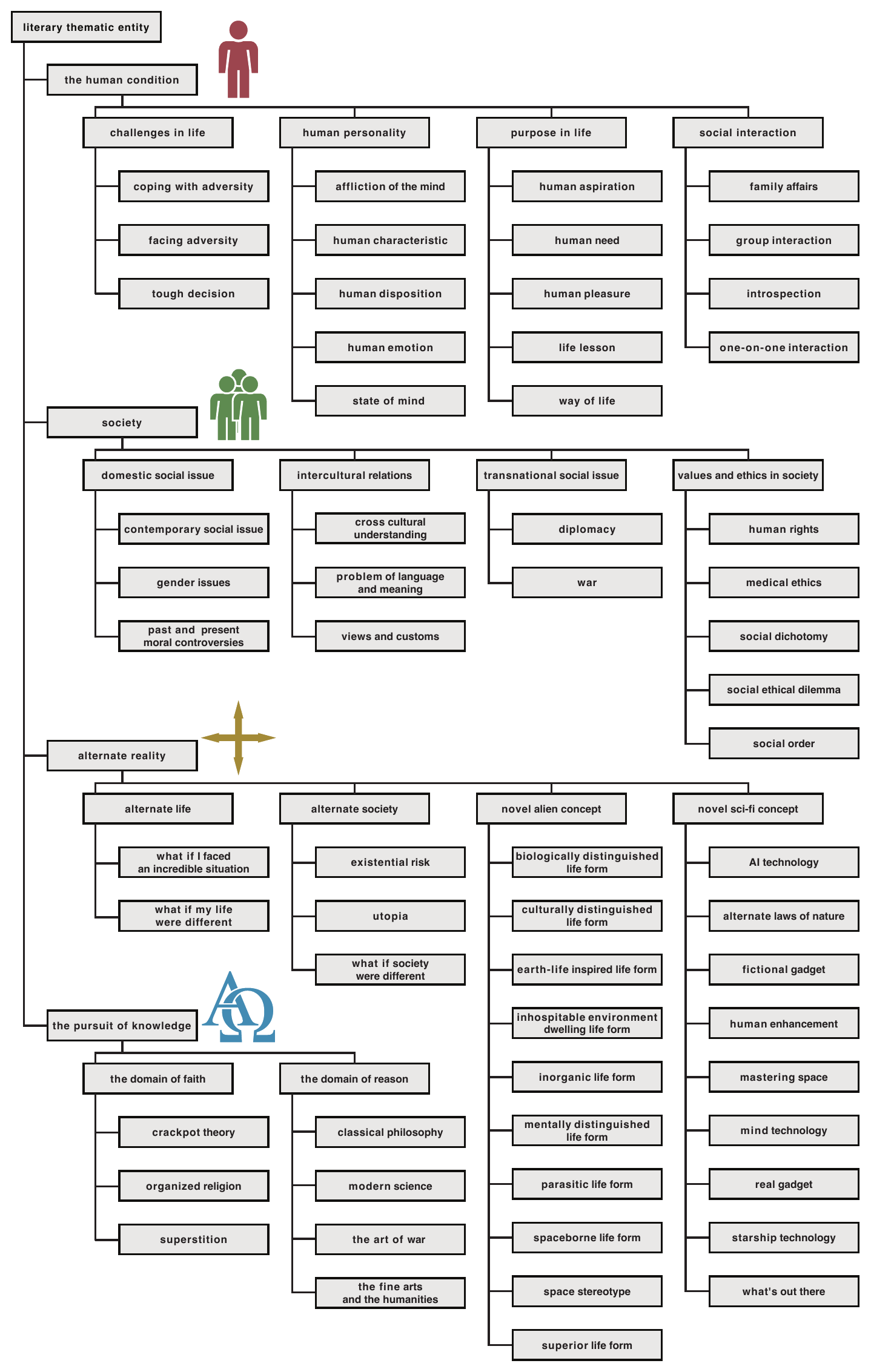}
\end{center}
\caption{Theme ontology domain hierarchies shown to three levels of depth.}\label{FIG_TO}
\end{figure}

A theme ontology is a controlled vocabulary of defined terms representing literary themes in fiction. In this section, we introduce a draft theme ontology covering all TOS, TAS, and TNG Star Trek television series episodes, of which there are~$280$. The ontology consists of~$1535$ unique themes arranged into the following four domains:
\begin{description}[before={\renewcommand\makelabel[1]{\bfseries ##1}}]
  \item[] \textbf{The Human Condition [\thinspace \tikz[baseline=-0.6ex]\draw[myred,fill=myred!50] (0,0) circle (1ex);\thinspace]:} Themes pertaining to ``characteristics, key events, and situations which compose the essentials of human existence, such as birth, growth, emotionality, aspiration, conflict, and mortality''~\footnote{Wiktionary (2018). URL: \url{https://en.wiktionary.org/wiki/human_condition} [Online; accessed 1-May-2018]}.
  \item[] \textbf{Society [\thinspace \tikz[baseline=-0.6ex]\draw[mygreen,fill=mygreen!50] (0,0) circle (1ex);\thinspace]:} Themes pertaining to a ``community of people living in a particular country or region and having shared customs, laws, and organizations''~\footnote{Oxford Dictionaries, Oxford University Press (2018). URL: \url{https://en.oxforddictionaries.com/definition/society}. [Online; accessed 1-May-2018]}.
  \item[] \textbf{The Pursuit of Knowledge [\thinspace \tikz[baseline=-0.6ex]\draw[myblue,fill=myblue!50] (0,0) circle (1ex);\thinspace]:} Themes pertaining to ``facts, information, and skills acquired through experience or education; the theoretical or practical understanding of a subject''~\footnote{Oxford Dictionaries, Oxford University Press (2018). URL: \url{https://en.oxforddictionaries.com/definition/knowledge}. [Online; accessed 1-May-2018]}.
  \item[] \textbf{Alternate Reality [\thinspace \tikz[baseline=-0.6ex]\draw[myyellow,fill=myyellow!50] (0,0) circle (1ex);\thinspace]:} Themes related to subject matter falling outside of reality as it is presently understood. These are classical science fiction and fantasy themes~\footnote{The Encyclopedia of Science Fiction (2018). URL: \url{http://www.sf-encyclopedia.com/}. [Online; accessed 1-June-2018]}.
\end{description}
Figure~\ref{FIG_TO} shows a bird's eye view of the ontology. The abstract theme ``literary thematic entity'' is taken as root theme. Each domain is structured as a tree descended from the root with ``the human condition'', ``society'', ``the pursuit of knowledge'', and ``alternate reality'' serving as the top themes of their respective domains. Each child theme is made to bear a subtype relationship with its parent. In the figure, the ontology tree structure is depicted to a height of three levels, although in reality it branches out a number levels further still, as summarized along with other information in Table~\ref{TAB_TO_Summary_Stats}.

In designing the ontology, we strive to define sibling themes so as to be mutually exclusive, but not necessarily jointly exhaustive. All non-root themes are accompanied by short definitions in an effort to make their range of applicability plain. We appeal to the principle of refutability as an anodyne to vagueness in definition writing. In other words, an acceptably defined theme will be such that there is the possibility of appealing to the definition to show that the associated theme is not featured in a story. Take ``to tell the truth vs. offering a comforting lie'' as an example, which is defined as ``A character must choose between telling a comforting white lie on one hand, and being honest on the other.''. We contend that definition writing of this sort helps to bring the conversation of whether a theme is featured in a given story into the realm of rational argumentation. The theme ontology dataset has been made accessible in a structured manner through the \proglang{R} package \pkg{stoRy}~\citep{Sheridan2017}. This paper uses version 0.1.1 of the ontology, which can be accessed through the analogously numbered \pkg{stoRy} package 0.1.1. Functions for exploring the ontology are described in the package reference manual. For example, the command \code{theme\$print()} prints summary information for the theme object \code{theme}, and the function \code{print\_tree} takes a theme object as input and prints the corresponding theme together with its descendants in tree format  to the console. We encourage non-R-users to explore the latest version of the ontology on the Theme Ontology website~\footnote{Theme Ontology (2018). URL: \url{http://www.themeontology.org}. [Online; accessed 1-May-2018]}. Previous versions of the ontology have been made available for download there.

It is crucial to bear in mind that the theme ontology is wholly separable from the Star Trek television series franchise. Many of the themes populating the ontology are sufficiently universal in scope to be helpful in the recording of themes for of almost any work of fiction. That said, we availed ourselves of these themes in the recording of themes for $280$ Star Trek television series episodes. We assigned central and peripheral themes for each episode. Note that centrality and peripherality are to be understood not as properties of a theme, but rather as relations between a theme and a story. Thus, it is perfectly fine for a given theme to be central to one story and peripheral to another. A theme is said to be \emph{observed} in a story when it is featured in such a way that none of its descendent themes are also featured. An ancestor of an observed theme in a story is said to be a \emph{latent} theme. Table~\ref{TAB_ST_Summary_Stats} shows a basic statistical summary of the thematic data by series. Summary statistics for observed themes in are shown in the upper part of the table, and observed and latent themes in the lower part.

Take the \emph{TOS} episode \emph{The Devil in the Dark} (1967) as an example. In the story, the starship USS Enterprise is dispatched to investigate rumors of a subterranean creature that is thought to be responsible for the destruction of equipment and the deaths of fifty men on the Janus~VI mining colony. Captain James T. Kirk and First Officer Spock discover a hideous ``silicon-based life form'' inhabiting the surrounding bedrock. The mother Horta, as the creature comes to be known, is an ``endangered species''\thinspace --\thinspace the last of its kind. Kirk is faced with a ``tough decision'' as the creature seemingly blocks the miners path to wealth: either commit genocide or forego plundering the mother Horta's natural resources. But Spock fortuitously manages to achieve a ``cross cultural understanding'' with the creature by means of a Vulcan mind-meld. An unsettling compromise is reached when the mother Horta agrees to help the miners locate ore deposits in the rock in exchange for a cessation of hostilities. The attentive viewer will note,  from the creature's perspective, an ``attack from outer space by a powerful conquering alien race'' (a.k.a. ``alien invasion'') with overtones of ``the morality of colonization'' by the episode's conclusion. This summarizes some of the more salient central story themes, as recorded by the authors. The proverb ``beauty is in the eye of the beholder'' is a noteworthy peripheral theme. Indeed, Spock learned, in the course of the mind-meld, that the human form is just as repellant to the mother Horta, as her appearance is to humans.

We tagged each of the $280$ episodes of TOS, TAS, and TNG with themes in a similar manner. The process according to which we assigned themes can be summed up as follows. We independently tagged episodes with themes and then compared notes with a view toward building a consensus set of themes for each episode. We aimed to abide in the principle of low-hanging fruit in the compilation of consensus themes. In the present context, this means we tried to ensure that at least the most salient topics featured in the episodes are covered by appropriate themes. Another principle guiding our thought process is the minimization of false positives (i.e., the tagging of episodes with themes that are not featured) at the expense of tolerating false negatives (i.e., neglecting to tag episodes with themes that they feature). This strategy amounts to erring on the side of caution. We fully acknowledge that this process lacks safeguards against the tagging of stories with themes that are idiosyncratic and unique to our point of view. We will return to this subject in the discussion section.

\section{Theme Enrichment Analysis} \label{SEC_TEA}

\begin{figure}[h!]
\begin{center}
\includegraphics[width=14.6cm]{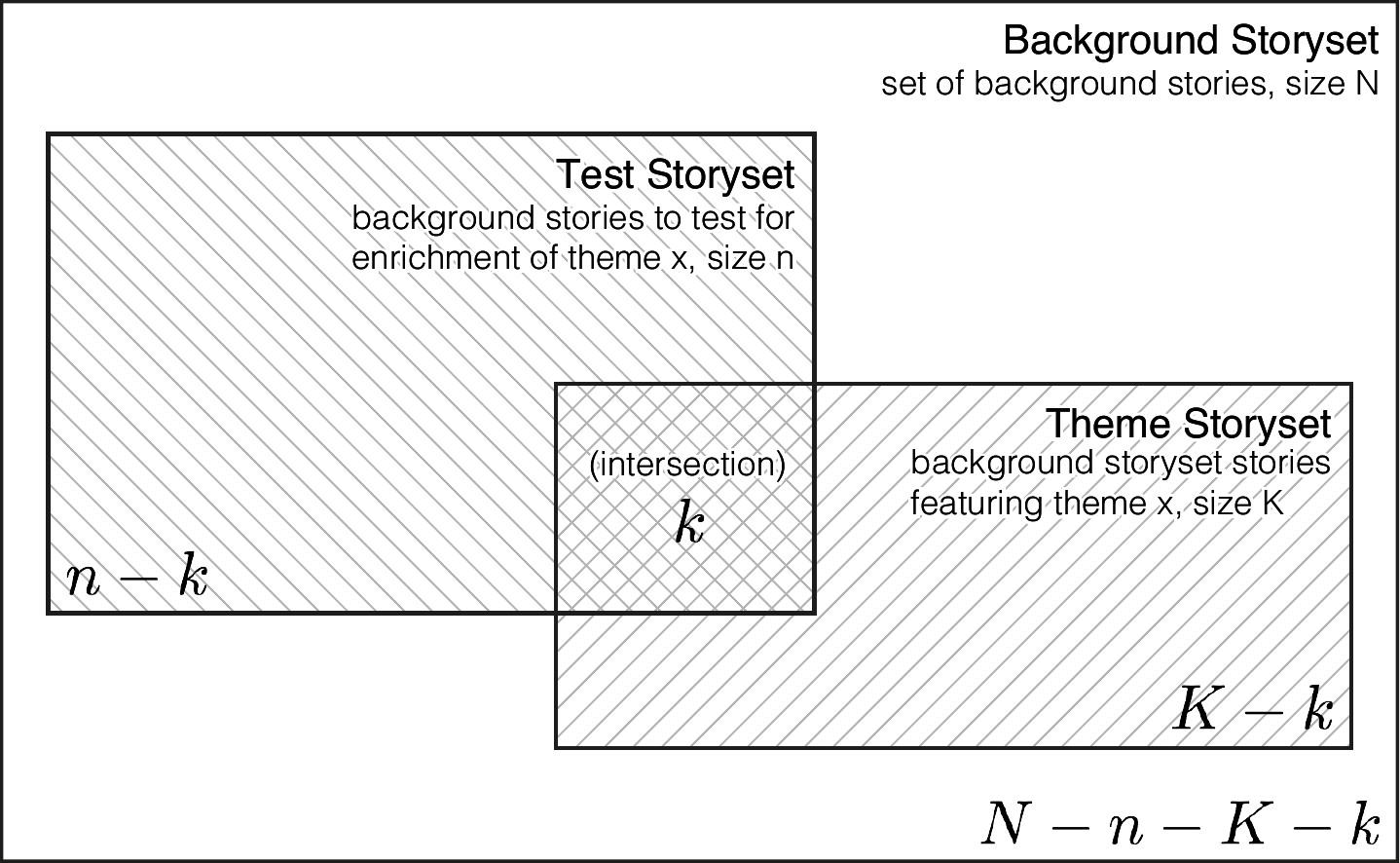}
\end{center}
\caption{Hypergeometric testing framework overview for testing whether a given theme $x$ is enriched in a test storyset relative to a background storyset. The various storysets and variables $n$, $k$, $K$, and $N$ are defined in the main text.}\label{FIG_TEA_Venn_Diagram}
\end{figure}

This section is devoted to an exposition on the hypergeometric testing approach to theme enrichment analysis.~The test uses the \emph{p}-value obtained from the hypergeometric cumulative distribution
\begin{equation*}~\label{EQ_Hypergeometric_Pvalue}
P(k, n, K, N) = \sum_{i = k}^{n} \frac{\binom{K}{i} \binom{N-K}{n-i}}{\binom{N}{n}}
\end{equation*}
to answer the question of whether a given theme occurs in a test storyset at a frequency significantly greater than would be expected by chance alone. In the equation, $n$ is the size of the test storyset, $k$ is the number of stories in the test storyset featuring the theme, $N$ is the size of the background storyset, and $K$ is the number of stories in the background storyset featuring the theme. The $K$ background stories featuring the theme determine what we will call the \emph{theme storyset}. Figure~\ref{FIG_TEA_Venn_Diagram} depicts the testing framework in Venn diagrammatic form. The value of $P(k, n, K, N)$ is the probability of observing at least $k$ stories featuring a given theme in a test storyset of size $n$ that is composed of stories drawn at random, without replacement, from the background storyset. As we explained in the introduction, a theme is deemed to be enriched in a test storyset with respect to a background storyset if its \emph{p}-value is less than a preselected significance level,~alpha. Whether the test storyset is found to be enriched for a given theme or not will necessarily depend on the choice of background storyset.

The hypergeometric testing procedure is silent on the matter of theme, test storyset, and background storyset selection. It falls on the investigator to first pose an interesting question, and then choose the hypergeometric test input storysets accordingly. In the previous section, we proposed that the \emph{TOS} episode \emph{The Devil in the Dark} (1967) featured an ``alien invasion''. Indeed, the theme has long been a favorite among sci-fi writers. More generally, the sci-fi genre has proven to be particularly well-suited to the exploration of ``existential risks'' (i.e., manners by which civilization on a planetary scale and beyond could come into jeopardy), of which ``alien invasion'' is but one, albeit far-fetched, example. It is easy to imagine a fictitious investigator who wishes to assess whether ``existential risk'' (theme) is enriched in \emph{TOS} (test storyset) with respect to all of \emph{TOS}, \emph{TAS}, and \emph{TNG} (background storyset). This casts the question ``Does \emph{TOS} stand out among Gene Roddenberry produced \emph{Star Trek} television series for its featuring of existential risks?'' in precise enough terms to be made amenable to theme enrichment analysis. As we will see in Section~\ref{SEC_TEA_Series}, the test results in a \emph{p}-value of $0.0002$, so that we may conclude the theme ``existential risk'' is enriched in \emph{TOS} at significance level alpha equal to $0.05$.

\begin{table}[h!]
\caption{Theme ontology domain hierarchy summary statistics.}
{\label{TAB_TO_Summary_Stats}}
\centering
\begin{tabular}{L{3cm}C{2.2cm}R{2.2cm}R{2.2cm}R{2.2cm}}
\toprule
\textbf{Domain root theme} & \textbf{Domain color-code} & \textbf{Theme count} & \textbf{Leaf theme count} & \textbf{Tree height} \\ \midrule
 the human condition & \tikz\draw[myred,fill=myred!50] (0,0) circle (1ex); & $631$ & $578$ & $6$ \\
 society & \tikz\draw[mygreen,fill=mygreen!50] (0,0) circle (1ex); & $279$ & $256$ & $4$ \\
 the pursuit of knowledge & \tikz\draw[myblue,fill=myblue!50] (0,0) circle (1ex); & $235$ & $217$ & $4$ \\
 alternate reality & \tikz\draw[myyellow,fill=myyellow!50] (0,0) circle (1ex); & $390$ & $357$ & $4$ \\ \bottomrule
\end{tabular}
\end{table}

In practice, the curious investigator will often wish to check whether each theme in an ontology is significantly enriched in a given test storyset relative to some background storyset. This raises the specter of multiple comparisons~\citep{Noble2009, Meijer2016, McDonald2014}. To illustrate the matter, recall that the \emph{p}-value in a hypothesis test is the probability of getting a result at least as extreme as the observed one, assuming the null hypothesis is true. A \emph{p}-value calculated to be less than the investigator's desired significance level, alpha, is interpreted as evidence in favor of the alternative hypothesis. The value of alpha sets an upper limit on the chance of making a false positive discovery that the investigator is willing to tolerate. Consider, for example, the classroom problem of testing whether a given coin is fair or biased. The conventional null and alternative hypotheses are that the coin is fair and biased, respectively. Choosing an alpha of $0.05$ translates into a $1$ in $20$ chance of concluding that the hypothetical coin in question is biased when it is actually fair (i.e., the null hypothesis is rejected when it is true). In the case when a large number hypothesis tests are conducted, chance dictates that some definite proportion of the \emph{p}-values obtained will be less than alpha, even when the truth of all the null hypotheses is assured. For example, suppose $1535$ fair coins are individually tested in the above manner. This is as many coins as there are themes in our ontology. If all~$1535$ coins are tested at an alpha of $0.05$, then the expected number of fair coins mistakenly identified as biased ones works out to be $0.05 \times 1535 \approx 77$. The aim of multiple comparison correction procedures is to limit the number of false positives. The Bonferroni correction and the false discovery rate~\citep{Benjamini1995, Benjamini2001} are the two most common procedures. However, it is inappropriate to correct for multiple comparisons in the context of testing for multiple enriched themes. This is because a given theme is either over-represented in a test storyset relative to a background or it is not. If a theme is significantly enriched at a given significance level, then it is a true positive by definition. Therefore it would be misguided to apply the sorts of standard correction procedures that are used in biology in the context of testing list of genes for enriched terms. For this reason, we refrain from correcting the \emph{p}-values calculated in this paper for multiple comparisons.

\begin{table}[h!]
\caption{Star Trek television franchise series episode theme summary statistics.}
{\label{TAB_ST_Summary_Stats}}
\centering
\begin{tabular}{L{0.5cm}R{1.5cm}R{4cm}R{4cm}}
\toprule
\textbf{Series} & \textbf{Episodes} & \shortstack[r]{\textbf{Mean number of central} \\ \textbf{themes per episode $\mathbf{\pm}$ s.d.}} & \shortstack[r]{\textbf{Mean number of peripheral} \\ \textbf{themes per episode $\mathbf{\pm}$ s.d.}} \\ \midrule
\multicolumn{4}{c}{Observed Themes} \\ \midrule
\emph{TOS} & $80$ & $12.54 \pm 4.40$ & $20.04 \pm 6.28$ \\
 \emph{TAS} & $22$ & $6.68 \pm 2.55$ & $3.50 \pm 2.39$ \\
 \emph{TNG} & $178$ & $11.72 \pm 4.42$ & $14.93 \pm 5.66$ \\ \midrule
 \multicolumn{4}{c}{Observed and Latent Themes} \\ \midrule
 \emph{TOS} & $80$ & $22.38 \pm 5.61$ & $29.38 \pm 6.21$ \\
 \emph{TAS} & $22$ & $7.55 \pm 2.06$ & $5.23 \pm 3.05$ \\
 \emph{TNG} & $178$ & $26.39 \pm 7.99$ & $30.72 \pm 8.96$ \\ \bottomrule
\end{tabular}
\end{table}

\section{A Study of Enriched Themes in Star Trek}\label{SEC_TEA_Application}

In this section, we present two cases studies of theme enrichment analysis applied to the Star Trek television franchise. As we explained in the introduction, the first examines how the Klingons changed from a tyrannical and expansionist empire in \emph{TOS} to an inward looking warrior culture in \emph{TNG}. The second, also described in the introduction, takes an in-depth look into significantly enriched themes by series. In short, we find that \emph{TOS} stands out for its treatment of what constitutes a good society and how one is to lead a good life within it, \emph{TAS} for novel alien and sci-fi concepts appropriate to an animated series, and \emph{TNG} for its comparatively refined treatment of the human condition. The message we wish to convey based on these case studies is that theme enrichment analysis is helpful when it comes to both the summarization what makes a storyset unique and the formulation of speculative hypotheses. Note that we include both central and peripheral themes in the analyses. In addition, we present the reassuring results of some negative control experiments, and compare the TF-IDF approach to enrichment with the results of our case studies. The theme enrichment analyses are easily replicated using version $0.1.1$ of the \pkg{stoRy} package.

\subsection{A Tale of Two Klingons}\label{SEC_TEA_Klingon}

The Klingons are an \"uber-belligerent humanoid species in the Star Trek alien pantheon. In \emph{TOS}, the Klingon Empire is made to pursue a harsh imperialist foreign policy, characterized by the unapologetic use of military force in the subjugation of their weaker neighbors. The Federation, by contrast, is portrayed as a group of confederated alien races, united under the common principles of human rights, equality, and interstellar cooperation. In the series, the Federation fights to check Klingon expansion in the galaxy. The conflict between the Federation and the Klingon Empire in \emph{TOS} is commonly understood as an allegory for the Cold War~\citep{Cantor2000}. According to this interpretation, the Federation represents the Western powers, and the Klingon Empire represents the Soviet Union; both as seen from a more or less contemporaneous American point of view. But by \emph{TNG}, Klingon society had undergone a radical transformation. The Klingons changed from being an avowed Federation enemy hell-bent on galactic domination to a loose Federation ally with a warrior culture preoccupied with internal struggle and the maintaining of cherished traditions in a changing world~\citep{Knight1998}. The changing face Klingon society has been examined in much detail in books~\citep{Taylor2002, Gonzalez2015, Telotte2008}, academic papers~\citep{Knight1998, Cantor2000}, and scattered online sources~\footnote{BBC (2011). Klingons and Commies. URL: \url{http://bbc.adactio.com/cult/st/original/commies.shtml}. [Online; posted 7-February-2011]}~\footnote{Roblin, S. (2016). Star Trek's Original Series Brought the Cold War Into Space. URL: \url{https://medium.com/war-is-boring/star-treks-original-series-brought-the-cold-war-into-space-91ae5a9291f4}. [Online; posted 28-October-2016]}~\footnote{Bondurant, T. (2016). From Kor to Discovery, We Say Qapla to 50 Years of Klingons. URL: \url{http://www.cbr.com/star-trek-klingons-50-anniversary/}. [Online; posted 19-February-2017]}.

In this case study, we use the Klingons as a positive control to demonstrate that our approach to theme enrichment analysis is able to distinguish the imperialist Klingons of \emph{TOS} from the warrior culture ones of \emph{TNG}. To this end, we curated a storyset consisting of all $26$ Klingon-centric \emph{TOS}, \emph{TAS}, and \emph{TNG} episodes; see Supplementary Note~2 for a complete episode listing. The criterion for episode inclusion is that the Klingons were deemed by the authors to have been featured throughout the episode in a way that is central to the story. We performed two theme enrichment analyses: 1) the test storyset of \emph{TOS}/\emph{TAS} Klingon-centric episodes against a background of all \emph{TOS}/\emph{TAS} episodes, 2) the test storyset of \emph{TNG} Klingon-centric episodes against a background of all \emph{TNG} episodes. The \emph{TOS} and \emph{TAS} episodes we pooled because \emph{TAS} is conventionally considered to be a continuation of \emph{TOS}. The \emph{TOS}/\emph{TAS} and  \emph{TNG} test storysets consist of $n=8$ and $n=18$ episodes, respectively. In each experiment, we calculated an enrichment score, that is, a hypergeometric test \emph{p}-value, for each of the $1535$ themes present in the ontology.

The test identified $27$ and $22$ enriched themes at significance level $0.05$ in the \emph{TOS}/\emph{TAS} and \emph{TNG} Klingon storysets, respectively. Tables containing the full results from these analyses are included in Supplementary Information File~2. Table~\ref{TAB_Klingons} shows the top $20$ most enriched themes for each analysis. We focus on interpreting the enriched themes shown in the table, but we note that the excluded ones are consistent with the interpretation we advance. With that caveat aside, an inspection of Table~\ref{TAB_Klingons} shows that our positive control test results are interpretable in a manner that is in keeping with expectation. Consider first those themes from the society domain. \emph{TOS}/\emph{TAS} Klingon ``imperialistic society'' posed a serious military threat to the Federation. When an inevitable ``transnational social issue'' flared up between the Federation and Klingon Empire, such as a ``conflict over a shared resource'', the resolution usually came about by either ``diplomatic negotiating'' or outright ``war''. Although in rare instances, Klingons and Federation members came to a ``cross cultural understanding'' when united by a common enemy. But \emph{TOS}/\emph{TAS} Klingon society had at its heart a ``conflict of moral codes'' with Federation ideology that proved insuperable. Where \emph{TOS}/\emph{TAS} Klingon society is enterprising and enthusiastic in its convictions, \emph{TNG} Klingon society is inward looking and gloomy. The society themes enriched in \emph{TNG} Klingon episodes pertain to internal conflicts, as evidenced by the themes ``racism in society'', ``religious fanaticism'', and ``war of succession''. No longer is the Klingon Empire striving to impose Klingon values on the galaxy by means of military force, but rather it is focused on its own internal affairs. This brings us to the human condition. In \emph{TOS}/\emph{TAS}, the human condition domain themes are almost all virtues possessed by the aliens that the Klingons conquered (i.e., ``pacifism'', ``humility'', ``patience'', and ``temperance''). On the other hand, the human condition domain themes enriched in \emph{TNG} (i.e., ``honor'', ``rage'', and ``loyalty'') are all signature \emph{TNG} Klingon characteristics. Notice, by contrast, that ``honor'' is nowhere to be found among the top $20$ significantly enriched \emph{TOS}/\emph{TAS} Klingon themes. In fact, it occurs as the $101$st ranked theme with a \emph{p}-value of $0.219$. What is more, in the one episode in which it was featured\thinspace --\thinspace \emph{Friday's Child} (1967)\thinspace --\thinspace it pertained not to the Klingons, but to their enemy the Capellans. A number of the human condition themes pertain to Worf trying to maintain his Klingon culture in a Human world\footnote{Worf, a main character in the \emph{TNG} series, is the lone Klingon crew member aboard the Enterprise-D starship.}. The themes ``the need for cultural heritage'' and ``belonging'' are two noteworthy examples. Other human condition themes surround aspects of life in a cut-throat warrior culture, like ``the lust for power'' and ``the desire for redemption''. Finally, it is pleasant to note that Klingon ``\"uber-belligerence''  and a passion for ``the art of war'' shine through in both cases.

\subsection{A Tale of Three Series}\label{SEC_TEA_Series}

\begin{figure}[!h]
\begin{center}
\includegraphics[width=14.6cm]{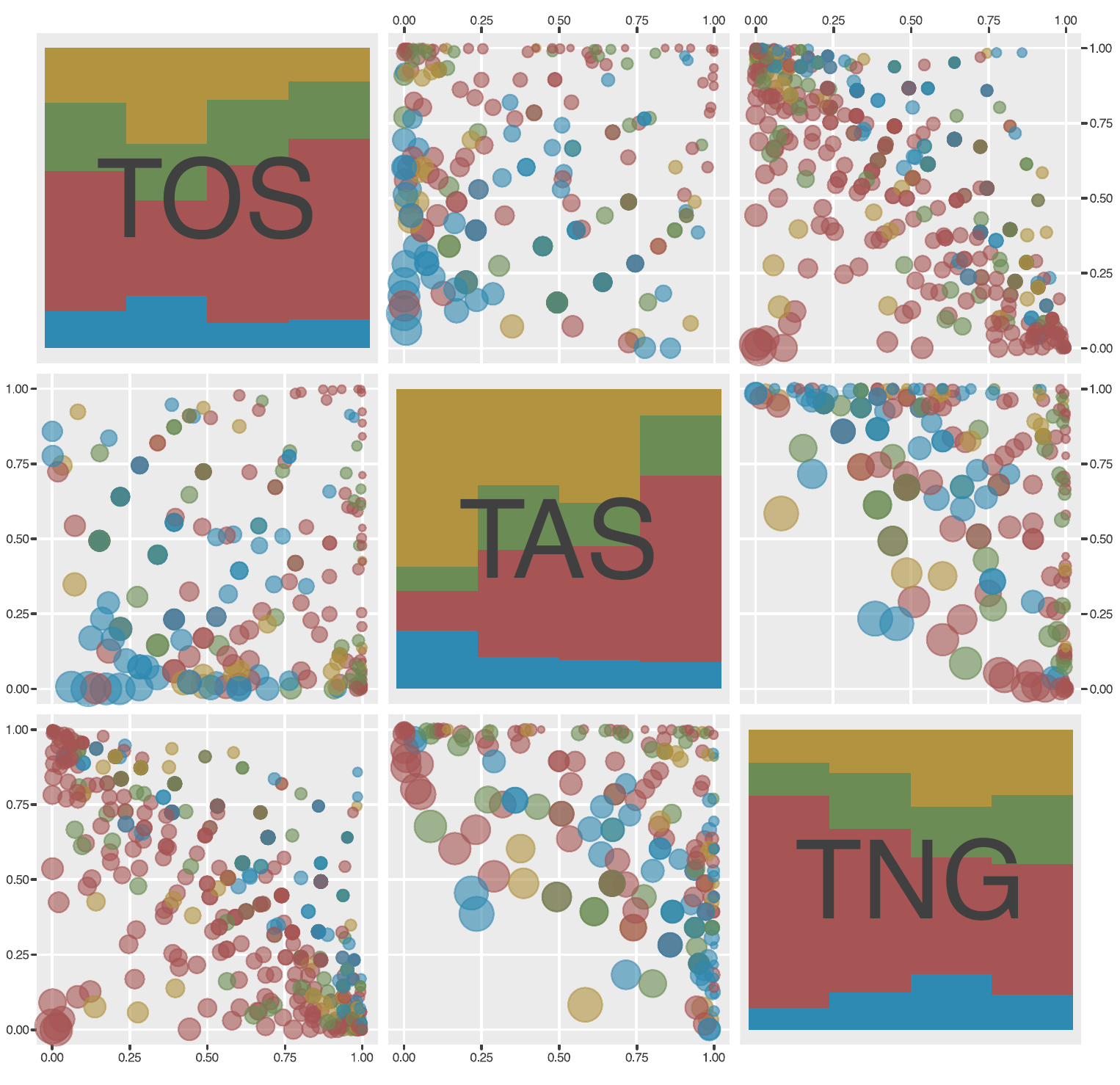}
\end{center}
\caption{Star Trek television series theme enrichment scatterplot matrix. Theme enrichment \emph{p}-value stacked percentage bar plots for \emph{TOS}, \emph{TAS}, and \emph{TNG} are plotted along the diagonal from top left to bottom right. For each series, the \emph{p}-values are divided into quartiles from left to right, and each stacked bar indicated the percentage of themes from each domain. Theme enrichment \emph{p}-value scores are plotted against one another for each pair of series in the off-diagonal panels. Circle size is proportional to the square root of the negative of the logarithm of the product of \emph{p}-values pairs. Color corresponds to theme domain: the human condition (red), society (green), the pursuit of knowledge (blue), alternate reality (yellow).\label{FIG_Series_Scatterplot_Matrix}}
\end{figure}

We used the test to identify enriched themes in \emph{TOS} ($120$ themes), \emph{TAS} ($6$ themes), and \emph{TNG} ($46$ themes) at significance level $0.05$. The background storyset in each case consists of the episodes from all the series combined. Here we report the outcomes of the analyses and show how they can aid in the generation of speculative hypotheses. In keeping with the Klingon case study from the previous subsection, we frame our hypotheses about \emph{TOS} and \emph{TNG} in terms of the top $20$ most enriched themes for each respective series. This, we contend, is enough to convey the merits of our methodology without burdening the reader with protracted syntheses of long lists of enriched themes, however pleasant an exercise the formulation of such syntheses might be for the authors. On the other hand, we take some leeway and extend our analysis of \emph{TAS} to the top $10$ most enriched themes. But we note that our general conclusions would remain unchanged had we limited our interpretations to the $6$ enriched themes at significance level $0.05$. Tables containing the full results from these analyses are included in Supplementary Information File 2. Star Trek enthusiasts will find few surprises in the kinds of themes that are shown to distinguish the respective series. To the layperson, however, the results of Table~\ref{TAB_Series} may be unexpected and serve as a useful point of departure for exploring the series. The stacked percentage bar plots of Figure~\ref{FIG_Series_Scatterplot_Matrix} show a broad pattern of human condition domain themes being enriched in \emph{TNG}, alternate reality domain themes in \emph{TAS}, and society domain themes in \emph{TOS} to some degree. The associated matrix scatterplot hints at some interesting enriched theme domain correlations between series. But let us proceed to inspect and compare more specific themes in order to gain a more nuanced understanding of the series.

\emph{TOS}: The two society domain themes ``female stereotype'' and ``gender issues'' stand out as they relate to the role of women in $1960$s society. The former is indicative of what Karen Blair and R.~P.~M. have described as a tendency in \emph{TOS} to portray females in such a manner as to ``affirm traditional male fantasies in a most direct and unenlightened way''~\citep{Blair1983}. The latter, however, is in keeping with another line of scholarly thought that contends \emph{TOS} made some positive contributions to the advancement of women in society~\citep{Ferguson1997, Vettel-Becker2014}. Three of the seven enriched alternate reality domain themes (i.e., ``alternate society'', ``existential risk'', and ``man-made existential risk'') confront viewers with ideas about how society could be changed for better or worse. In particular, the emphasis on existential risks is likely a reflection of the Cold War and relatively fresh memories of the Second World War~\citep{Cantor2000}. In light of this, it is interesting to note that five of the most enriched human condition domain themes (i.e., ``wrath'', ``facing a fight to the death'', ``rage'', ``unpleasant emotion'', and ``disagreeable characteristic'') are also closely tied to conflict. The remaining human condition domain themes (i.e., ``way of life'', ``purpose in life'', ``personal ethical dilemma'', ``tough decision'', and ``the need for a challenge in life'') pertain to life choices and decision making. Speculating about why this might be a feature of \emph{TOS} relative to the later series is left to the reader. Suffice it to say that they, like all the most enriched human condition domain themes in \emph{TOS}, are notably different from the most enriched themes in \emph{TNG}.

\emph{TAS}: Seven of the ten most enriched themes can be labeled simply as fanciful notions. Four are typical sci-fi themes: ``earth-life inspired life form'', ``life-support belt'', ``miscellaneous life form'', and ``what if my life were different''. The remaining two themes ``Chariots of the Gods'' and ``Atlantis'' refer to the ``crackpot theories'' that aliens made contact with humans in ancient times and Atlantis was a civilization with advanced technology, respectively. That such themes are enriched in \emph{TAS} can be explained by the fact that it is the only animated series of the trio. This would have released the authors from constraints otherwise imposed by the need for costly props and special effects (see Table~\ref{TAB_Series} for some examples) and allowed them to further unleash their imagination. We hypothesize the lack of emotion-related themes may be partially explained by early animation technology's inability to approach the nuances of facial expression and body language that the consummate actors of \emph{TOS} and \emph{TNG}, such as William Shatner and Patrick Stewart, would routinely employ.

\emph{TNG}: Nearly all of the $20$ most enriched \emph{TNG} themes relate to individual human experience. About half of them are descendants of the themes ``family affairs'': ``familial love'', ``growing up'', ``mother and son'', ``maternal love'', ``adolescence'', ``familial relations'', ``father and son'', ``paternal love'', and ``child rearing''. The others are not dissimilar: ``human emotion'', ``heavenly virtue'', ``human personality'', ``social interaction'', ``pride'', ``belonging'', and ``introspection''. Anyone familiar with the android Data will recognize that the themes ``android'' and ``AI point of view'' relate to stories about individual human experience as well. One theme that stands out for not falling under the human condition is ``virtual reality room'', a particular sci-fi concept that refers to the holodeck in \emph{TNG}, which has become something of a meme in its own right. Why it is that \emph{TNG} so distinctly features these family affair, relationship, and emotional themes is of course open to interpretation. We speculatively hypothesize that an inexorable trend in modern television has been towards vapid character development designed to evoke safe familiarity rather than intellectual stimulus or moral controversy. Be that as it may, it is safe to say that the main characters in \emph{TNG} have more elaborate background stories, subtle personality traits, and complicated interpersonal relationships than the main characters of the two earlier series.

\subsection{Negative Control Experiments}\label{SEC_TEA_Negative_Control}

We performed a series of negative control experiments. In one such experiment, we performed enrichment analyses for $1000$ test storysets consisting of $n=8$ randomly selected \emph{TOS}/\emph{TAS} episodes against the same background. This mirrors the \emph{TOS}/\emph{TAS} Klingon positive control settings. The average number of significantly enriched themes was $10 \pm 5$ at significance level $\alpha = 0.05$. The corresponding Klingon storyset has $28$ significantly enriched themes at the same significance level. This is noticeably more enriched themes on average than would be expected by chance. In a similar negative control with $n=18$ for the test storyset, we found the average number of significantly enriched themes to be $13 \pm 5$ relative to a background of all \emph{TNG} episodes. The corresponding Klingon storyset has $22$ significantly enriched themes. Again this is more than would be expected to be enriched by chance.

\subsection{Hypergeometric Test and TF-IDF Comparison}\label{SEC_TEA_Comparison}

TF-IDF is a statistic that is used in data mining to measure the importance of a word in a document in a collection of documents~\citep{Rajaraman2015}. It is implemented in such textual analysis tools as the Stanford Topic Modeling Toolbox~\citep{Ramage2009}, TAPoR~\footnote{TAPoR 3 Discover research tools for studying texts (2018). URL: \url{http://tapor.ca/tools}. [Online; accessed 1-May-2018]}, TOME~\citep{Klein2015}, Word Seer~\citep{Muralidharan2013}, and Voyant Tools~\footnote{Sinclair, S., Rockwell, G. and the Voyant Tools Team, Voyant Tools (2012). URL: \url{https://voyant-tools.org/}. [Online; accessed 1-May-2018]}. In this subsection, we take a TF-IDF approach to the identification of enriched themes in the Klingon and series case studies, and compare the results with those obtained using the hypergeometric test. We implemented the TF-IDF formula $- k/n \times \log(K/N)$, where $k/n$ is the term frequency, and $-\log(K/N)$ is the logarithmically scaled inverse document frequency. Table~S2 contains the top $20$ TF-IDF scoring themes in \emph{TOS}/\emph{TAS} Klingon and \emph{TNG} Klingon episodes. Table~S3 contains the analogous results for the individual series \emph{TOS}, \emph{TAS}, and \emph{TNG}. For Tables~S2-S3, refer to Supplementary Note 3. The results significantly overlap with those obtained using the hypergeometric test. For the Klingons of \emph{TOS}/\emph{TAS} a remarkable $20$ out of the top $20$ themes are held in common (it turns out to be $38$ out of the top $50$), and for the Klingons of \emph{TNG} the figure amounts to $14$ out of $20$. The numbers for the individual series \emph{TOS}, \emph{TAS}, and \emph{TNG} are $12/20$, $6/10$, and $8/20$, respectively. What is more, the scatterplots of Figure~S2 show broad correlation between TF-IDF scores and hypergeometric test logarithmically scaled \emph{p}-values. What we take from this is that TF-IDF and the hypergeometric test approaches agree in their broad outlines, but differ in the details. Based on these outcomes, we contend that elucidating the mathematical relationship between TF-IDF and the hypergeometric test would make for an interesting future work. For the present, we merely point out that one advantage of the hypergeometric test is that the p-values obtained therefrom are, on the face of it, more amenable to interpretation than TF-IDF scores. To be specific, the p-value of a theme is interpretable as the probability of drawing a sample consisting of at least $k$ stories that feature the theme in $n$ draws, without replacement, from a background of $N$ stories of which exactly $K$ feature the theme. By contrast, the search for a simple probabilistic interpretation of TF-IDF scores remains a subject of ongoing research~\citep{SparckJones1988, Robertson2004, Havrlant2017}.

\begin{table}[!h]
\caption{Enriched themes in Klingon-centric episodes relative to \emph{TOS}/\emph{TAS} and \emph{TNG} backgrounds, respectively.}\label{TAB_Klingons}
\centering
\begin{tabular}{R{1cm}L{3cm}C{1cm}R{1cm}R{1.5cm}L{6cm}}
\toprule
\textbf{Rank} & \textbf{Theme} & \textbf{Domain} & $\boldsymbol{k/K}$  & \textbf{P value} & \textbf{Comment} \\ \midrule
\multicolumn{6}{c}{Top 20 Enriched Themes in \emph{TOS}/\emph{TAS} Klingon Episodes} \\ \midrule
1 & \"uber-belligerent alien & \tikz\draw[myyellow,fill=myyellow!50] (0,0) circle (1ex); & $5/5$ & $<0.0001$ & The Klingons are belligerence personified. \\
2 & diplomatic negotiating & \tikz\draw[mygreen,fill=mygreen!50] (0,0) circle (1ex); & $4/7$ & $0.0005$ & Teetered on brink of war with the Federation.\\
3 & culturally distinguished life form & \tikz\draw[myyellow,fill=myyellow!50] (0,0) circle (1ex); & $6/20$ & $0.0006$& Parent theme of \"uber-belligerent alien. \\
4 & man vs. beast & \tikz\draw[myred,fill=myred!50] (0,0) circle (1ex); & $3/5$ & $0.0030$ & Beasts faced: mugato (x1), tribble (x2).\\
5 & diplomacy & \tikz\draw[mygreen,fill=mygreen!50] (0,0) circle (1ex); & $5/19$ & $0.0053$ & Parent theme of diplomatic negotiating. \\
6 & conflict over a shared resource & \tikz\draw[mygreen,fill=mygreen!50] (0,0) circle (1ex); & $2/2$ & $0.0054$ & Vied with Federation over galactic resources. \\
7 & atrocities of war & \tikz\draw[mygreen,fill=mygreen!50] (0,0) circle (1ex); & $2/2$ & $0.0054$ & Not above committing war crimes. \\
8 & tribble & \tikz\draw[myyellow,fill=myyellow!50] (0,0) circle (1ex); & $2/2$ & $0.0054$ & Mortal enemy of the Klingons. \\
9 & pacifism & \tikz\draw[myred,fill=myred!50] (0,0) circle (1ex); & $3/7$ & $0.0098$ & Pacifists made easy targets for conquest. \\
10 & military tactics & \tikz\draw[myblue,fill=myblue!50] (0,0) circle (1ex); & $3/7$ & $0.0098$ & Masters of the art of war. \\
11 & war & \tikz\draw[mygreen,fill=mygreen!50] (0,0) circle (1ex); & $5/22$ & $0.0110$ & Revelled in warfare and conquest. \\
12 & transnational social issue & \tikz\draw[mygreen,fill=mygreen!50] (0,0) circle (1ex); & $6/32$ & $0.0111$ & Parent theme of diplomacy and war. \\
13 & the art of war & \tikz\draw[myblue,fill=myblue!50] (0,0) circle (1ex); & $5/23$ & $0.0136$ & Parent theme of military tactics. \\
14 & miscellaneous life form & \tikz\draw[myyellow,fill=myyellow!50] (0,0) circle (1ex); & $3/8$ & $0.0150$ & Parent theme of mugato and tribble. \\
15 & imperialistic society & \tikz\draw[mygreen,fill=mygreen!50] (0,0) circle (1ex); & $3/8$ & $0.0150$ & Klingon Empire subjugated weaker neighbors. \\
16 & conflict of moral codes & \tikz\draw[mygreen,fill=mygreen!50] (0,0) circle (1ex); & $2/3$ & $0.0157$ & Klingon imperialism vs. Federation benevolence. \\
17 & cross cultural understanding & \tikz\draw[mygreen,fill=mygreen!50] (0,0) circle (1ex); & $4/16$ & $0.0198$ & Parent theme of conflict of moral codes. \\
18 & humility & \tikz\draw[myred,fill=myred!50] (0,0) circle (1ex); & $3/9$ & $0.0217$ & Exhibited in the face of Klingon aggression. \\
19 & patience & \tikz\draw[myred,fill=myred!50] (0,0) circle (1ex); & $3/10$ & $0.0297$ & Exhibited in the face of Klingon aggression. \\
20 & temperance & \tikz\draw[myred,fill=myred!50] (0,0) circle (1ex); & $2/4$ & $0.0301$ & In contrast with Klingon licentiousness. \\ \midrule
\multicolumn{6}{c}{Top 20 Enriched Themes in \emph{TNG} Klingon Episodes} \\ \midrule
1 & \"uber-belligerent alien & \tikz\draw[myyellow,fill=myyellow!50] (0,0) circle (1ex); & $14/14$ & $<0.0001$ & The Klingons are still belligerence personified. \\
2 & honor & \tikz\draw[myred,fill=myred!50] (0,0) circle (1ex); & $13/18$ & $<0.0001$ & Central to the TNG Klingon way of life. \\
3 & culturally distinguished life form & \tikz\draw[myyellow,fill=myyellow!50] (0,0) circle (1ex); & $14/53$ & $<0.0001$ & Parent theme of \"uber-belligerent alien. \\
4 & the need for cultural heritage & \tikz\draw[myred,fill=myred!50] (0,0) circle (1ex); & $5/8$ & $0.0003$ & Worf cherished his Klingon cultural heritage. \\
5 & rage & \tikz\draw[myred,fill=myred!50] (0,0) circle (1ex); & $7/23$ & $0.0032$ & Flying into a violent rage is a key Klingon trait. \\
6 & racism in society & \tikz\draw[mygreen,fill=mygreen!50] (0,0) circle (1ex); & $3/5$ & $0.0080$ & Worf despised the Romulans down to the last man. \\
7 & facing wrongful accusations & \tikz\draw[myred,fill=myred!50] (0,0) circle (1ex); & $3/5$ & $0.0080$ & E.g. Worf falsely accused of treason.\\
8 & belonging & \tikz\draw[myred,fill=myred!50] (0,0) circle (1ex); & $7/27$ & $0.0090$ & Worf was the lone Klingon in Starfleet. \\
9 & religious fanaticism & \tikz\draw[mygreen,fill=mygreen!50] (0,0) circle (1ex); & $2/2$ & $0.0100$ & E.g. Klingon religious fanatics cloned their messiah. \\
10 & war of succession & \tikz\draw[mygreen,fill=mygreen!50] (0,0) circle (1ex); & $2/2$ & $0.0100$ & Klingon civil war. \\
11 & father and son & \tikz\draw[myred,fill=myred!50] (0,0) circle (1ex); & $6/21$ & $0.0101$ & E.g. Worf and his son Alexander. \\
12 & guilt and evidence & \tikz\draw[mygreen,fill=mygreen!50] (0,0) circle (1ex); & $4/11$ & $0.0163$ & Applies by and large to non-Klingon characters. \\
13 & the art of war & \tikz\draw[myblue,fill=myblue!50] (0,0) circle (1ex); & $6/24$ & $0.0207$ & Still masters of the art of war. \\
14 & the lust for power & \tikz\draw[myred,fill=myred!50] (0,0) circle (1ex); & $3/7$ & $0.0244$ & Various Klingons lusted to rule Klingon Empire. \\
15 & the desire for redemption & \tikz\draw[myred,fill=myred!50] (0,0) circle (1ex); & $2/3$ & $0.0280$ & Worf sought to redeem father's soiled reputation. \\
16 & loyalty & \tikz\draw[myred,fill=myred!50] (0,0) circle (1ex); & $7/33$ & $0.0295$ & A ruthless yet loyal alien race. \\
17 & agreeable characteristic & \tikz\draw[myred,fill=myred!50] (0,0) circle (1ex); & $14/96$ & $0.0305$ & Parent theme of honor. \\
18 & brother and brother & \tikz\draw[myred,fill=myred!50] (0,0) circle (1ex); & $3/8$ & $0.0366$ & Worf and his brother Kurn. \\
19 & cooperation & \tikz\draw[myred,fill=myred!50] (0,0) circle (1ex); & $5/20$ & $0.0370$ & Klingon Empire and Fed. sometimes joined forces.\\ 
20 & surprise & \tikz\draw[myred,fill=myred!50] (0,0) circle (1ex); & $6/27$ & $0.0377$ & E.g. Worf was surprised to find he had a young son. \\ \bottomrule
\end{tabular}
\end{table}

\begin{table}[!h]
\caption{Enriched themes for each Star Trek television series relative to a \emph{TOS}/\emph{TAS}/\emph{TNG} background.}\label{TAB_Series}
\centering
\begin{tabular}{R{1cm}L{3cm}C{1cm}R{1cm}R{1.5cm}L{6cm}}
\toprule
\textbf{Rank} & \textbf{Theme} & \textbf{Domain} & $\boldsymbol{k/K}$  & \textbf{P value} & \textbf{Comment} \\ \midrule
\multicolumn{6}{c}{Top 20 TOS Enriched Themes} \\ \midrule
1 & female stereotype & \tikz\draw[mygreen,fill=mygreen!50] (0,0) circle (1ex); & $19/24$ & $<0.0001$ & Reinforced some outmoded female stereotypes. \\
2 & wrath & \tikz\draw[myred,fill=myred!50] (0,0) circle (1ex); & $41/85$ & $<0.0001$ & Violent outbursts abounded. \\
3 & alternate society & \tikz\draw[myyellow,fill=myyellow!50] (0,0) circle (1ex); & $45/100$ & $<0.0001$ & Mind-openingly different societies explored. \\
4 & facing a fight to the death & \tikz\draw[myred,fill=myred!50] (0,0) circle (1ex); & $13/16$ & $<0.0001$ & Captain Kirk no stranger to such altercations. \\
5 & what if I had to fight to the death & \tikz\draw[myyellow,fill=myyellow!50] (0,0) circle (1ex); & $13/16$ & $<0.0001$ & Captain Kirk won ten and lost one to Spock. \\
6 & gender issues & \tikz\draw[mygreen,fill=mygreen!50] (0,0) circle (1ex); & $26/46$ & $<0.0001$ & Challenged some outmoded sexist attitudes, too. \\
7 & rage & \tikz\draw[myred,fill=myred!50] (0,0) circle (1ex); & $27/50$ & $<0.0001$ & Hurled bowls of Vulcan plomeek soup and so on. \\
8 & real gadget & \tikz\draw[myyellow,fill=myyellow!50] (0,0) circle (1ex); & $8/8$ & $<0.0001$ & Flatscreen TV, lie detector, teleconferencing, etc. \\
9 & unpleasant emotion & \tikz\draw[myred,fill=myred!50] (0,0) circle (1ex); & $78/237$ & $<0.0001$ & Parent theme of rage. \\
10 & alternate life & \tikz\draw[myyellow,fill=myyellow!50] (0,0) circle (1ex); & $39/87$ & $<0.0001$ & Fantastical things befalling people explored. \\
11 & way of life & \tikz\draw[myred,fill=myred!50] (0,0) circle (1ex); & $24/45$ & $0.0001$ & Ideologies on how to lead a good life explored. \\
12 & purpose in life & \tikz\draw[myred,fill=myred!50] (0,0) circle (1ex); & $70/201$ & $0.0002$ & Parent theme of way of life. \\
13 & existential risk & \tikz\draw[myyellow,fill=myyellow!50] (0,0) circle (1ex); & $37/83$ & $0.0002$ & Various threats to human civilization explored. \\
14 & imperialistic society & \tikz\draw[mygreen,fill=mygreen!50] (0,0) circle (1ex); & $8/9$ & $0.0002$ & Federation a bastion against imperialism. \\
15 & what if I faced an incredible situation & \tikz\draw[myyellow,fill=myyellow!50] (0,0) circle (1ex); & $30/64$ & $0.0003$ & Examples of the sorts of incredible situations faced are provided in the main text. \\
16 & disagreeable characteristic & \tikz\draw[myred,fill=myred!50] (0,0) circle (1ex); & $48/121$ & $0.0003$ & E.g. complacency, deviousness, rudeness, etc.\\
17 & personal ethical dilemma & \tikz\draw[myred,fill=myred!50] (0,0) circle (1ex); & $45/112$ & $0.0005$ & Characters faced with difficult moral choices. \\
18 & tough decision & \tikz\draw[myred,fill=myred!50] (0,0) circle (1ex); & $61/169$ & $0.0005$ & Parent of personal ethical dilemma. \\
19 & man-made existential risk & \tikz\draw[myyellow,fill=myyellow!50] (0,0) circle (1ex); & $22/43$ & $0.0006$ & E.g. the dangers of WMDs and societal laziness. \\
20 & the need for a challenge in life & \tikz\draw[myred,fill=myred!50] (0,0) circle (1ex); & $7/8$ & $0.0008$ & Fundamental component of the \emph{TOS} ethos. \\ \midrule
\multicolumn{6}{c}{Top 10 TAS Enriched Themes} \\ \midrule
1 & earth-life inspired life form & \tikz\draw[myyellow,fill=myyellow!50] (0,0) circle (1ex); & $6/15$ & $0.0004$ & Avians, felinoids, insectoids, slug-like aliens, etc. \\
2 & life-support belt & \tikz\draw[myyellow,fill=myyellow!50] (0,0) circle (1ex); & $3/3$ & $0.0004$ & A belt-like device that functions as a spacesuit.\\
3 & miscellaneous life form & \tikz\draw[myyellow,fill=myyellow!50] (0,0) circle (1ex); & $5/13$ & $0.0017$ & E.g. an alien composed of autonomous parts. \\
4 & man vs. beast & \tikz\draw[myred,fill=myred!50] (0,0) circle (1ex); & $3/8$ & $0.0187$ & Beasts faced: ie-matya, rock beast, tribble. \\
5 & Chariots of the Gods & \tikz\draw[myblue,fill=myblue!50] (0,0) circle (1ex); & $2/4$ & $0.0326$ & Aliens supplied the ancients with technology. \\
6 & genetic engineering & \tikz\draw[myblue,fill=myblue!50] (0,0) circle (1ex); & $2/4$ & $0.0326$ & Captain Kirk genetically engineered to have gills. \\
7 & what if my life were different & \tikz\draw[myyellow,fill=myyellow!50] (0,0) circle (1ex); & $6/38$ & $0.0615$ & E.g. what if I were a specimen animal in a zoo. \\
8 & crackpot theory & \tikz\draw[myblue,fill=myblue!50] (0,0) circle (1ex); & $5/30$ & $0.0726$ & Parent of Atlantis and Chariots of the Gods. \\
9 & man vs. nature & \tikz\draw[myred,fill=myred!50] (0,0) circle (1ex); & $3/13$ & $0.0732$ & Nature faced: arctic, desert, volcanic morass. \\
10 & Atlantis & \tikz\draw[myblue,fill=myblue!50] (0,0) circle (1ex); & $1/1$ & $0.0791$ & Aquan society submerged in watery cataclysm. \\ \midrule
\multicolumn{6}{c}{Top 20 TNG Enriched Themes} \\ \midrule
1 & human emotion & \tikz\draw[myred,fill=myred!50] (0,0) circle (1ex); & $176/265$ & $<0.0001$ & Virtues, vices, emotions pleasant and unpleasant. \\
2 & virtual reality room & \tikz\draw[myyellow,fill=myyellow!50] (0,0) circle (1ex); & $31/32$ & $<0.0001$ & I.e., the holodeck. \\
3 & familial love & \tikz\draw[myred,fill=myred!50] (0,0) circle (1ex); & $49/59$ & $0.0002$ & Bonds between family members emphasized. \\
4 & growing up & \tikz\draw[myred,fill=myred!50] (0,0) circle (1ex); & $40/47$ & $0.0003$ & Problems faced in early life treated. \\
5 & heavenly virtue & \tikz\draw[myred,fill=myred!50] (0,0) circle (1ex); & $105/144$ & $0.0004$ & Emphasized human virtues detailed in main text. \\
6 & android & \tikz\draw[myyellow,fill=myyellow!50] (0,0) circle (1ex); & $33/38$ & $0.0006$ & I.e., Lieutenant Commander Data. \\
7 & human personality & \tikz\draw[myred,fill=myred!50] (0,0) circle (1ex); & $176/271$ & $0.0008$ & Parent theme of human emotion. \\
8 & social interaction & \tikz\draw[myred,fill=myred!50] (0,0) circle (1ex); & $166/249$ & $0.0008$ & People interacting in groups explored. \\
9 & pride & \tikz\draw[myred,fill=myred!50] (0,0) circle (1ex); & $51/64$ & $0.0012$ & Just one of the many human vices examined. \\
10 & belonging & \tikz\draw[myred,fill=myred!50] (0,0) circle (1ex); & $27/31$ & $0.0020$ & Finding one's place in a group explored. \\
11 & mother and son & \tikz\draw[myred,fill=myred!50] (0,0) circle (1ex); & $24/27$ & $0.0021$ & E.g. Beverly Crusher and her son Wesley. \\
12 & introspection & \tikz\draw[myred,fill=myred!50] (0,0) circle (1ex); & $106/149$ & $0.0026$ & Characters keen on self-analysis of mental states.  \\
13 & maternal love & \tikz\draw[myred,fill=myred!50] (0,0) circle (1ex); & $20/22$ & $0.0029$ & A mother's love for her child featured. \\
14 & AI point of view & \tikz\draw[mygreen,fill=mygreen!50] (0,0) circle (1ex); & $26/30$ & $0.0030$ & The world as might be viewed by an AI shown. \\
15 & family affairs & \tikz\draw[myred,fill=myred!50] (0,0) circle (1ex); & $91/126$ & $0.0035$ & Ups and downs of family life examined. \\
16 & adolescence & \tikz\draw[myred,fill=myred!50] (0,0) circle (1ex); & $19/21$ & $0.0044$ & Difficulties faced by teenagers explored. \\
17 & familial relations & \tikz\draw[myred,fill=myred!50] (0,0) circle (1ex); & $81/111$ & $0.0044$ & Parent of mother and son/father and son. \\
18 & father and son & \tikz\draw[myred,fill=myred!50] (0,0) circle (1ex); & $21/24$ & $0.0065$ & E.g. Worf and his son Alexander Rozhenko. \\
19 & paternal love & \tikz\draw[myred,fill=myred!50] (0,0) circle (1ex); & $23/27$ & $0.0087$ & Parents' love for their children featured. \\
20 & child rearing & \tikz\draw[myred,fill=myred!50] (0,0) circle (1ex); & $10/10$ & $0.0094$ & Struggles of raising a child explored. \\ \bottomrule
\end{tabular}
\end{table}

\section{Discussion}\label{SEC_Discussion}

The primary aim of this paper has been to introduce the hypergoemetric test for theme enrichment analysis to the digital humanities community. We consider our proposed draft theme ontology and toy Star Trek thematic dataset to be contributions of a secondary nature. The hypergeometric testing approach to the identification of enriched themes in a list of stories equips the digital humanist with a new weapon to wield in their document analyses. We have demonstrated the potential of the hypergeometric test by applying it to Star Trek television franchise thematic data. In the first place, we found the lists of enriched themes produced by the test to be helpful for identifying what makes the corresponding storysets special and for generating speculative hypotheses. In the second place, we argued that the hypergeometric test compares favorably with the commonly used TF-IDF statistic when it comes to answering the question of whether such-and-such a theme (or any term for that matter) occurs in a storyset at a frequency significantly greater than would be expected by chance. But we would be remiss not to touch on the potential for ``garbage in, garbage out'' (GIGO) to bias and confound theme enrichment analyses. In other words, the result of a theme enrichment analysis using the hypergeometric test is at best as good as the themes on which it is based. The GIGO menace warrants special mention in the context of thematic analyses. The reason is that themes, unlike objective scientific entities such as genes, are generally complex abstractions of a highly subjective nature. We, therefore, stress the preliminary nature of our work in applying the hypergeometric test to thematic data. All that said, it is our hope that this work will contribute to the judicious use of the hypergeometric test in the digital humanities in the fullness of time.

Two main obstacles stand in the way of making theme enrichment analysis practical on a large-scale. First, a protocol for tagging stories with themes in such a manner that stories can be meaningfully compared in terms of their shared themes must be developed. We have taken a first step toward addressing this need by proposing a draft theme ontology. Moving forward, we aim to make the ontology compliant with Basic Formal Ontology design best practices~\citep{Arp2015}. This includes the incorporating of related ontologies such as the Emotion Ontology~\citep{Hastings2011} to name but one. Ontology design is an open-ended process, subject to setbacks and changes of direction. It is plain that our draft theme ontology will be no exception. However, we point out that even if the structure of the ontology changes markedly, many of the themes will remain intact as presently defined. Second, a large-scale database of compatibly themed stories is required. To this end, we have launched the Theme Ontology (beta version) online community platform~\footnote{Theme Ontology (2018). URL: \url{http://www.themeontology.org}. [Online; accessed 1-May-2018]}. The website features an ever-expanding controlled vocabulary of defined themes, hierarchically arranged into our draft theme ontology. Community members are encouraged to tag whatever stories (e.g., short stories, novels, films, TV shows, etc.) they please with themes drawn from the ontology, and adorn the ontology with newly coined themes as necessary. Stories are manually tagged with themes at present. Topic modeling techniques~\citep{Blei2012} as implemented in such software packages as MALLET~\footnote{McCallum, A. K., MALLET: A Machine Learning for Language Toolkit (2002). URL: \url{http://mallet.cs.umass.edu/}. [Online; accessed 1-May-2018]}, the Python module gensim~\citep{Rehurek2010}, topicmodels~\citep{Grun2011}, and the \proglang{R} package \pkg{tm}~\citep{Feinerer2008} have been used successfully to identify literary themes in text copora~\citep{Jockers2013a, Jockers2013b, Goldstone2014, Boyd-Graber2017}. In the future, we plan to use topic modeling to automatically collect themes for large numbers of stories in order to grow the Theme Ontology database. An interesting challenge awaits in figuring out how to adapt the current methods for automatic topic labeling~\citep{Lau2011, Basave2014, Bhatia2017} to the problem of mapping identified topics to themes in the ontology. Lastly, a theme enrichment analyzer web application is available for download at the Theme Ontology GitHub repository at~\url{https://github.com/theme-ontology/shiny-apps}. Tools from the the \pkg{stoRy} package, including our theme enrichment test, will be made accessible as web applications on the Theme Ontology website in order to help users analyze curated thematic datasets. It is our aim to build up a large-scale database of freely available themed stories that can be analyzed using web applications via this story theming system.

On the identification of themes in stories, something must be said. We emphasize that our tagging of Star Trek television series episodes with themes in this paper was manual, subjective to our point of view, and not fully replicable. In general, theme identification is admittedly subjective, but this is not to say the endeavor is altogether arbitrary. The Oxford Dictionaries definition of theme quoted in the introduction, which proves adequate for most literary-critical purposes, is problematic from the present point of view insofar as it designates a theme to be a property of a story. Instead, we propose to consider a theme as a relation between story and partaker thereof. According to this subjective view, it is entirely possible for partaker~A to contend that theme X is featured in story Y, but not partaker~B. Objectivity is approached to the extent that universal agreement among story partakers is attained. The potential for a system of this kind to degenerate into a wasteland of subjectivity depends on the extent to which themes can be made precise. At the Theme Ontology community platform, we are presently drafting a policies and guidelines document that will emphasize the need for clarity in and verifiability of theme definitions. The theme ``the desire for vengeance'', which is defined as ``A character seeks vengeance over a perceived injury or wrong.'', constitutes a model definition. Growing pains are inevitable. But by concentrating on cataloging precisely defined and verifiable themes, i.e., the low-hanging fruit, we hope to ensure that Theme Ontology becomes a useful literary studies resource in the future.

There are a few points to mention in closing. First, the toy thematic dataset we have introduced in the present work consists of the combined $280$ episodes of \emph{TOS}, \emph{TAS}, and \emph{TNG}. {We contend this sufficed for our purpose of demonstrating the value of theme enrichment analysis. But there entire Star Trek franchise is made up of $740$ episodes and films. A good future work would be to round out the toy dataset with themed episodes from all the other series along with the films. Second, the hypergeometric test is designed to answer the question: what themes in a test storyset of interest stand out against a background storyset. But other interesting literary questions are also amenable to statistical investigation. For example, the investigator may wish to discover a subset of stories in a storyset that have similar themes by performing a clustering analysis. Another example is time-series analysis for the study of how theme usage changes over time in a storyset with timestamped stories. In the future, the \pkg{stoRy} package should be extended to include statistical methods to address questions of these sorts. Lastly, computer-assisted qualitative data analysis software is sometimes used for document analysis in the social sciences. ATLAS.ti~\footnote{ATLAS.ti Scientific Software Development GmbH (1993-2017). ATLAS.ti.}, NVivo~\footnote{QSR International Pty Ltd. (2015). NVivo qualitative data analysis Software Version 11.}, and MAXQDA~\footnote{VERBI Software (1989-2017). MAXQDA, software for qualitative data analysis. are three such programs that allow users to manage, analyze, and visualize data related to text, audio, and video documents. It is possible to use these sorts of programs to annotate stories with themes and subsequently explore the correlations among them. However, to our knowledge, none of these programs implement the hypergeometric test. The same thing holds for the popular textual analysis tools the Stanford Topic Modeling Toolbox~\citep{Ramage2009}, TAPoR~\footnote{TAPoR 3 Discover research tools for studying texts (2018). URL: \url{http://tapor.ca/tools}. [Online; accessed 1-May-2018]}, TOME~\citep{Klein2015}, Word Seer~\citep{Muralidharan2013}, and Voyant Tools~\footnote{Sinclair, S., Rockwell, G. and the Voyant Tools Team, Voyant Tools (2012). URL: \url{https://voyant-tools.org/}. [Online; accessed 1-May-2018]}.} In the future, it could be profitable to augment these programs with a hypergeometric test function for term enrichment.

\section*{Conflict of Interest Statement}
The authors declare that the research was conducted in the absence of any commercial or financial relationships that could be construed as a potential conflict of interest.

\section*{Author Contributions}
MO and PS collected the data. MO and PS conceived the analysis. PS conducted the analysis. MO and PS wrote the manuscript.

\section*{Acknowledgments}
We kindly thank Adam Clay, Atsushi Niida, and Sergio Jimenez for their helpful comments and suggestions.

\bibliographystyle{frontiersinSCNS_ENG_HUMS} 
\bibliography{references}




\end{document}